\documentclass[11pt,a4paper]{article}
\usepackage{amsfonts,graphicx}

\newcommand{\G}{{\cal G}}
\newcommand{\Gr}{\Gamma}
\newcommand{\I}{\mathbb{I}}
\newcommand{\1}{\mbox{\sf id}}
\newcommand{\C}{\mathbb{C}}
\newcommand{\R}{\mathbb{R}}
\newcommand{\T}{\mathbb{T}}

\newcommand{\LL}{\mbox{L}_2}

\title{Spin filtering on a ring with Rashba hamiltonian}
\author{Mark Harmer\\
mark.harmer@maths.anu.edu.au\\
Department of Mathematics\\
Australian National University\\
Canberra 0200\\
Australia}

\begin{document}

\maketitle 

\begin{abstract}
We consider a quantum graph consisting of a ring with Rashba hamiltonian and an arbitrary number of  semi-infinite wires attached. We describe the scattering matrix for this system and investigate spin filtering for a three terminal device.
\end{abstract}

\section{Introduction}
Much work has been done on spin related transport properties of nanoelectronic devices resulting in interesting applications, for example the so called spin field effect transistor proposed by Datta and Das \cite{Dat:Das}. There has been particular interest in using the Rashba effect to manipulate the spin degree of freedom in such systems \cite{GBZS, Kis:Kim1, MMK, NMT, SGB, SGZ}. In this paper we model a simple system exhibiting the Rashba effect, viz. a ring with Rashba hamiltonian attached to an arbitrary number of `free' wires, using so called solvable models \cite{Alb:Kur, Pav, Har4, Kost:Sch2}. This means that we approximate the system by a---one dimensional---graph on which is defined an appropriate self adjoint Schr\"{o}dinger operator. The advantage of this approach is that, as the name suggests, it allows us to get explicit expressions for the scattering matrix, and hence for the transport properties of the system, in this case in terms of the Greens function of the ring and the boundary conditions at the vertices. \\
Our particular interest in considering this model is to investgate the possibility of constructing a spin filter. Various approaches have been taken to filter spin: we mention \cite{GBZS} in which the authors construct a spin filter using a four terminal device with the Rashba effect as well as \cite{Str:Seb} where the authors achieve spin filtering using a two terminal device and a magnetic field. A third approach, discussed in \cite{Kis:Kim0, Kis:Kim1}, uses a three terminal device with the Rashba effect and to some extent was the motivation for this paper. \\ 
It is known that a device with two terminals and time reversal symmetry cannot polarise spin currents \cite{Kis:Kim2} (the device in \cite{Str:Seb} does not have time reversal invariance due to the magnetic field). Nevertheless, Kiselev and Kim \cite{Kis:Kim0, Kis:Kim1} show that a three terminal device with time reversal symmetry and a particular geometric symmetry can make an effective spin filter. We consider the same geometry as considered in \cite{Kis:Kim1}, viz. a ring with three wires and symmetry with respect to reflection across the line defined by the `incoming' wire. Whereas Kiselev and Kim assume the Rashba effect is localised at the `incoming' terminal in our model the Rashba hamiltonian is present uniformly on the whole ring. Kiselev and Kim use a sophisticated numerical model of the system to calculate transport properties while our model is of course solvable. \\
We believe that the formalism of solvable models offers, in general, advantages over numerical studies in that it allows us to derive explicit expressions for scattering properties thereby identifying principal features of the system. Ideally, these may be used to help optimise the design (for instance for spin filtering). In particular, for the three terminal device described above we investigate how the polarisation is related to the resonant eigenvalues on the ring, the Rashba coefficient and the angle of attachment of the wires. We observe, as did Kiselev and Kim, that this system may be used as an efficient spin filter.
\section{Quantum graph with Rashba hamiltonian}
We consider a ring shaped quantum waveguide where the width of the waveguide and the incident electron energy is such that the ring may be considered one-dimensional. Furthermore, we assume that there is structural inversion asymmetry \cite{Win} so that a Rashba term appears in the hamiltonian on the ring. Normalising the radius to one it can be shown \cite{MMK, SGZ} that the hamiltonian has the form
$$
H_0 f = D^2_0 \, f - \left( \frac{\alpha}{2} \right)^2 f
$$
where
\begin{eqnarray*}
D_0 & = & -\frac{1}{i}\frac{d}{d\theta} + \frac{\alpha}{2} \sigma_r \, , \\
\sigma_r & = & \sigma_x\, \cos (\theta) + \sigma_y\, \sin (\theta) \, ,
\end{eqnarray*} 
$\theta\in [0,2\pi)$ is the local coordinate on the ring; $\sigma_{x}$, $\sigma_{y}$, $\sigma_{z}$,  $\1$ denote the Pauli spin matrices and the unit matrix respectively; and $\alpha$ describes the strength of the Rashba spin-orbit coupling. The solutions of the eigenequation, $H_0 f = k^2 f$, are
\begin{equation}\label{eigensol}
f_{\pm, 0} (\theta , k) = e^{-i\sigma_{z} \theta/2}\, e^{-i\sigma_{y} \varphi/2}\,
e^{\pm i\sigma_{z} \kappa_{\pm} \left( \theta - \pi \right)}
\end{equation}
where $\kappa_{\pm} = \sqrt{k^2+\frac{\alpha^2}{4}} \pm \sqrt{\frac{1}{4}+\frac{\alpha^2}{4}}$ and $\tan(\varphi) = \alpha$, $\varphi\in (-\frac{\pi}{2},\frac{\pi}{2})$. The eigenvalues on the ring 
$$
\lambda_{\pm,n} = n^2 - \left( {\textstyle \frac{1}{2}} \pm n \right) \left( \sqrt{1+\alpha^2} -1 \right) 
$$
correspond to the zeroes of $\cos (\kappa_{\pm} \pi)$. Each eigenvalue $\lambda_{\pm,n}$ has multiplicity two, the corresponding eigenspace is spanned by $\left\{ e^{\pm in\theta} \chi_{_\uparrow}, \, e^{\mp in\theta} \chi_{_\downarrow}\right\}$ where
$$
\chi_{_\uparrow} = \left( \begin{array}{c}
\cos (\varphi/2) \\
e^{i\theta} \sin (\varphi/2) 
\end{array} \right) \, , \quad
 \chi_{_\downarrow} = \left( \begin{array}{c}
- e^{-i\theta} \sin (\varphi/2)  \\
\cos (\varphi/2)
\end{array} \right) \, .
$$
Since $\lambda_{+,n}=\lambda_{-,-n}$ and $\lambda_{+,n}\le\lambda_{-,n}$ we assume $n\in \{1,2,\ldots\}$ for $\lambda_{+,n}$ and $n\in \{0,1,\ldots\}$ for $\lambda_{-,n}$. Finally, we note that the twofold degeneracy of the eigenvalues drops to a fourfold degeneracy when $\sqrt{1+\alpha^2} -1=m\in \{0,1,\ldots\}$. In this case we see that $\lambda_{-,n}=\lambda_{+,n+m}$. \\
Mostly we will write eigenfunctions with both spin eigenstates together in a $2\times 2$ matrix in order to simplify notation. In particular the solutions $f_{\pm, 0}$ may be used to find the Greens function, ie. the continuous solution of
\begin{eqnarray*}
H_0 \, G \left(\theta , \eta ; k^2\right) & = & k^2\, G (\theta , \eta ; k^2) \\
\left. \frac{\partial }{\partial \theta} G (\theta , \eta ; k^2 ) \right|^{\theta = \eta^{+}}_{\theta = \eta^{-}} & = & \1 \\
G (\theta , \eta ; k^2) & = & G^{\star} (\eta , \theta ; k^2) \, , \quad k\in\R\setminus \sigma\left( H_{\alpha} \right) \, ,
\end{eqnarray*}
which is in fact
\begin{eqnarray}
& & G \left(\theta , \eta ; k^2\right)  \label{Gfn} \\
& = & \left[  f_{+,0} (\theta) \frac{ e^{- i\sigma_{z} \kappa_{+} \eta} }{\cos (\kappa_{+} \pi )} - f_{-,0} (\theta) \frac{ e^{ i\sigma_{z} \kappa_{-} \eta} }{\cos (\kappa_{-} \pi )} \right] \frac{e^{-i\sigma_{y} \varphi/2} }{2 i (\kappa_{+} + \kappa_{-} )} e^{ i\sigma_{z}  \eta/2}\, \sigma_{z} \nonumber \\
& = & \frac{ e^{-i\sigma_{z} \theta /2}\, e^{-i\sigma_{y} \varphi/2} }{2 i (\kappa_{+} + \kappa_{-} )}
\left[  \frac{ e^{ i\sigma_{z} \kappa_{+} \left( \theta - \eta - \pi \right)} }{\cos (\kappa_{+} \pi )} - \frac{ e^{ -i\sigma_{z} \kappa_{-} \left( \theta - \eta - \pi \right)} }{\cos (\kappa_{-} \pi )} \right] e^{-i\sigma_{y} \varphi/2} e^{ i\sigma_{z}  \eta/2}\, \sigma_{z} \, . \nonumber
\end{eqnarray}
Here we take $\theta - \eta \in [0,2\pi)$. \\
We assume that the ring is attached to $n$ semi-infinite wires. On each wire we have a `free' hamiltonian
$$
H_j f_j = D_j^2 \, f_j \, , \qquad D_j = - \frac{1}{i}\frac{d}{dx_j} \, ,
$$
with generalised eigenfunctions
$$
f_{\pm, j} = e^{ \pm i\, \mbox{\small \1} k x_j }
$$
where $j\in\{1,\ldots ,n\}$ is the index for the wire and $x_j$ is the coordinate on the respective wire. \\
We write the hamiltonian on the whole system
$$
H = H_0 \oplus \sum^{n}_{j=1} H_j 
$$
and consider this as an operator on the Hilbert space $\LL (\Gr ,\C^2) = \LL (\T ,\C^2 )\oplus \sum^{n}_{j=1}\LL (\R_+ ,\C^2 )$ of spinor valued functions on the graph $\Gr$ consisting of the ring $\T$ with $n$ wires $\R_+$ attached. To define this as a self adjoint operator we need to correctly define the domain of $H$ which is related to self adjoint boundary conditions arising from the vanishing of the boundary form
\begin{eqnarray}
\left( H f , g \right) - \left( f , H g \right) & = & i \sum^n_{j=1} \left. \left( \langle D_j f , g \rangle + \langle f , D_j g \rangle \right) \right|_{x_j = 0} \nonumber \\
& & \mbox{} + i \sum^n_{j=1} \left. \left( \langle D_0 f , g \rangle + \langle f , D_0 g \rangle \right) \right|^{\theta = \theta^{+}_j}_{\theta = \theta^{-}_j} \, . \label{bform}
\end{eqnarray}
Generally these boundary conditions are parameterised by a unitary matrix, for details see \cite{Alb:Kur, Har4, Pav}. Here $\left( \cdot , \cdot \right)$ is the inner product on $\LL (\Gr ,\C^2)$, $\langle \cdot , \cdot \rangle$ is the inner product on spinors and $\{ \theta_j \}^n_{j=1}$ are the points where the wires are attached to the ring. \\
We always assume that each vertex has three incident edges, ie. no two wires are attached at the same point on the ring. For a given vertex we denote by $\left\{ \psi_i \right\}^{3}_{i=1}$ the values of the eigenspinor on edge $i$ in the limit as we approach the vertex and by $\left\{ \psi^{\prime}_i \right\}^{3}_{i=1}$ the values of the outward derivative on edge $i$ in the limit as we approach the vertex. In this paper we assume the following boundary conditions at the vertices
\begin{equation}\label{bndcnd}
\beta^{-1} \psi_1 = \psi_2 = \psi_3 \, , \quad \beta\psi^{\prime}_{1} + \psi^{\prime}_{2} + \psi^{\prime}_{3} = 0 \, ,
\end{equation}
motivated by the fact that they are closely related to the ansatz for the scattering matrix of the T-junction as described in the physics literature \cite{Datt} (we describe this relationship in the first appendix). Here edge $i=1$ is the semi-infinite wire while edges $i=2,3$ are on the ring. The coefficient $\beta$ describes the strength of the coupling between the wire and the ring. \\
These boundary conditions (\ref{bndcnd}) are assumed to hold with the same $\beta$ for each component of the spinor, ie. the coupling is independent of spin. However, $\beta$ may in general be different at each vertex or point where a wire is attached to the ring. It is clear that these boundary conditions are self-adjoint, ie. (\ref{bform}) vanishes (see \cite{Kost:Sch2} for a discussion of boundary conditions in the presence of magnetic terms). \\
\section{Scattered waves and the scattering matrix}
The scattered waves $\psi_{i}$ are eigenfunctions on the quantum graph satisfying the boundary conditions at the vertices and having the following form on the wires:
\begin{equation}\label{SWwires}
\psi_{i} = f_{+,i} +  f_{-,i} S_{ii} \oplus \sum _{j\ne i} f_{-,j} S_{ji} \, .
\end{equation}
We reiterate that $\psi_{i}$ corresponds to two spinor valued waves, one with spin up and one with spin down incident waves. Similarly, the components of the scattering matrix $S_{ji}$ , $i,j \in \{1,\ldots , n \}$, are $2\times 2$ matrix valued. Due to the nature of the boundary conditions we can define the scattered wave on the ring as
\begin{equation}\label{SWring}
\psi_{i} =  \sum _{k} G_{k} A_{ki} 
\end{equation}
where $G_{k} = G \left(\theta, \theta_k \right)$, $\{\theta_i\}^n_{i=1}$ are the points where the wires are attached to the ring and $A$ is a matrix of coefficients. \\
The boundary conditions for the scattered waves can be neatly expressed using the Greens function. Defining
$$
\G_{jk} = G \left(\theta_j , \theta_k ; k^2\right) 
$$
we see from (\ref{SWwires}) that on the wires
$$
\left. \psi_{i} \right|_j = \delta_{ji} + S_{ji} \, , \quad
\left. \psi^{\prime}_{i} \right|_j = ik \left( \delta_{ji} - S_{ji} \right) 
$$
where $\left. \cdot \right|_{j}$ denotes evaluation at the point where the $j$-th wire is attached to the ring and $\delta_{ji}$ should be interpreted as matrix with matrix valued components. Likewise from (\ref{SWring}) we have
$$
\left. \psi_{i} \right|_j = \sum_{k} \G_{jk}A_{ki}   \, , \quad
\left. \psi^{\prime}_{i} \right|^{\theta = \theta^{+}_{j}}_{\theta = \theta^{-}_{j}} = A_{ji} 
$$
on the ring where we have used the property of the derivative of the Greens function. 
It is then easy to see that the boundary conditions in (\ref{bndcnd}) can be written
$$
\I + S = \beta\, \G A \, , \quad ik \beta \left(\I - S\right) = -A
$$
respectively where $\beta$ is a diagonal matrix containing the coupling strengths for the $n$ vertices. Solving for the scattering matrix we get
\begin{equation}\label{Smatrix}
S = \left( ik\beta\G\beta + \I\right)  \left( ik\beta\G\beta - \I\right)^{-1} \, .
\end{equation}
This result can be generalised, at least in principle, using the techniques described in \cite{Har5, Kost:Sch1} to find the scattering matrix of a quantum graph consisting of $n$ semi-infinite wires attached to a compact graph consisting of $m$ rings with Rashba term connected by edges of finite length. \\
\section{Spin filtering using a three terminal Rashba ring}
\begin{figure}[ht]\hspace*{-35mm}
\includegraphics{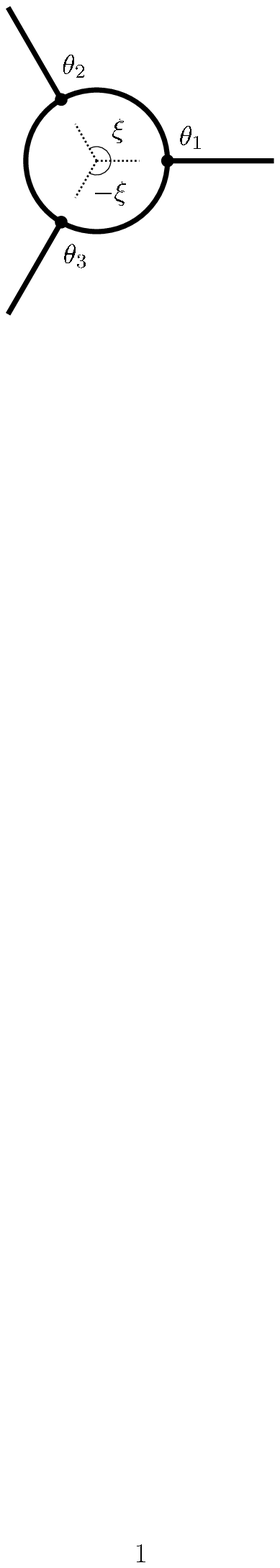}
\vspace*{-210mm}
\caption{The three terminal Rashba ring.}\label{sfil}
\end{figure}

\begin{figure}[ht]
\hspace*{-10mm}
\includegraphics[height=15cm,width=14cm]{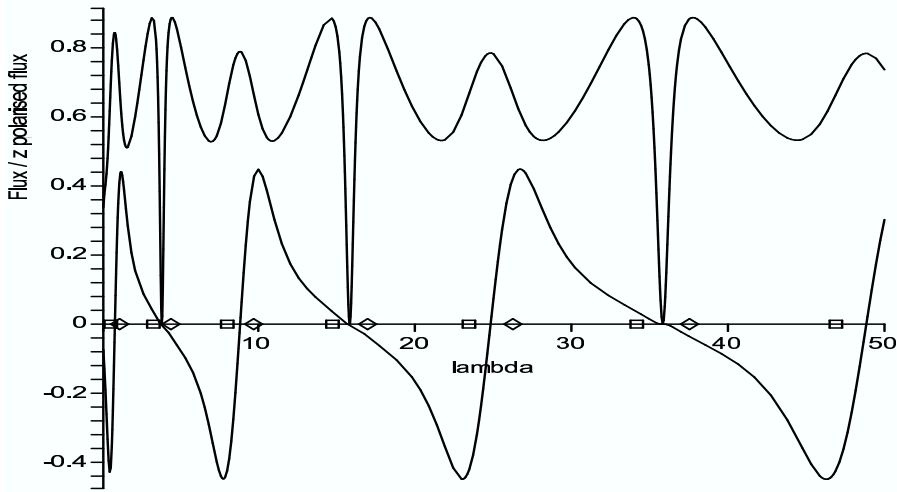}\vspace*{-7.6cm}
\caption{$T_{21}$ and $P_{21,z}$ for $\alpha=0.8$ and $\xi=\pi/2$.}\label{pi2}
\end{figure}

\begin{figure}[ht]
\begin{center}
\vspace*{-1.1cm}
\includegraphics[height=15cm,width=14cm]{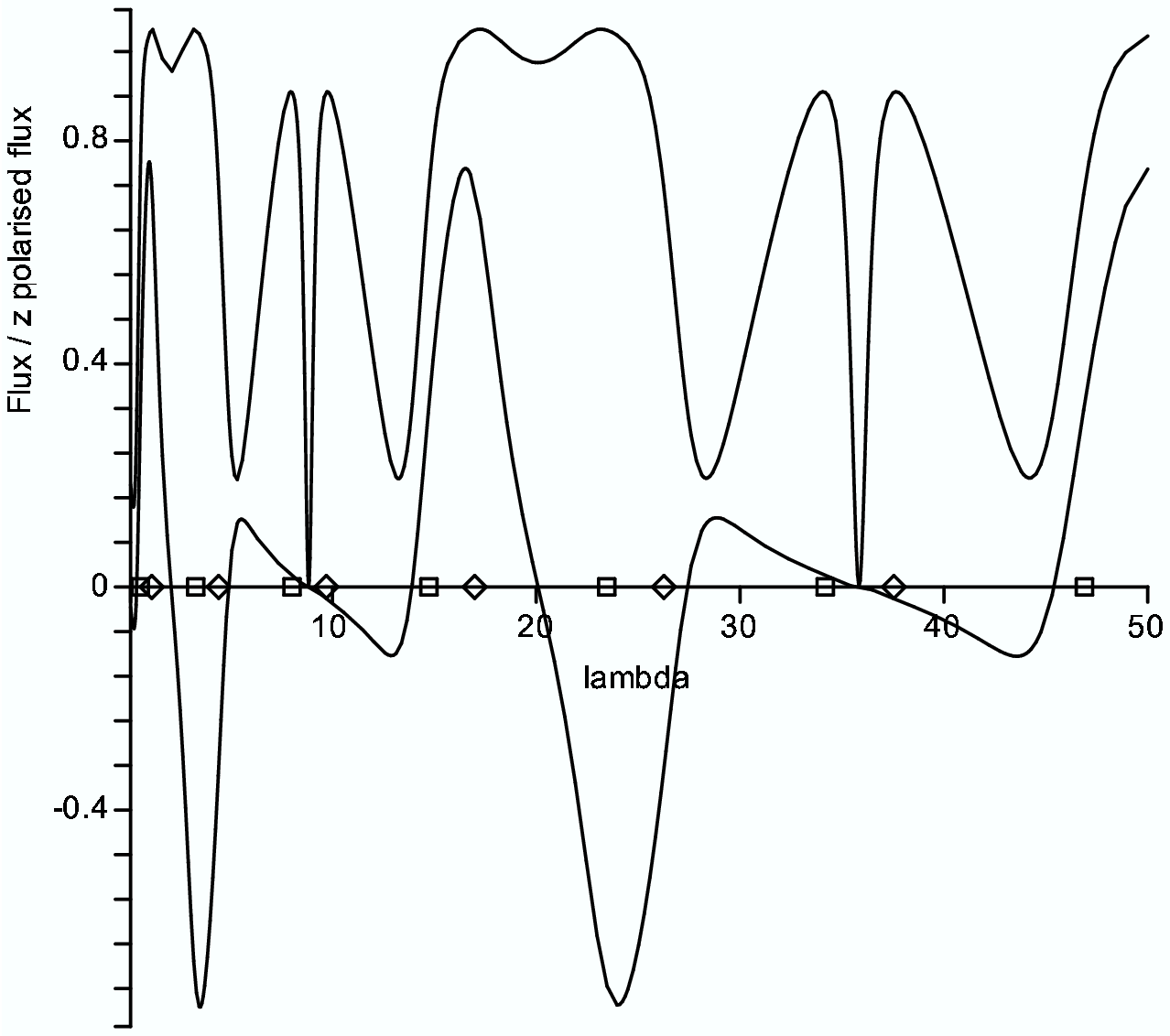}\vspace*{-7.6cm}
\caption{$T_{21}$ and $P_{21,z}$ for $\alpha=0.8$ and $\xi=\pi/3$.}\label{pi3}
\includegraphics[height=15cm,width=14cm]{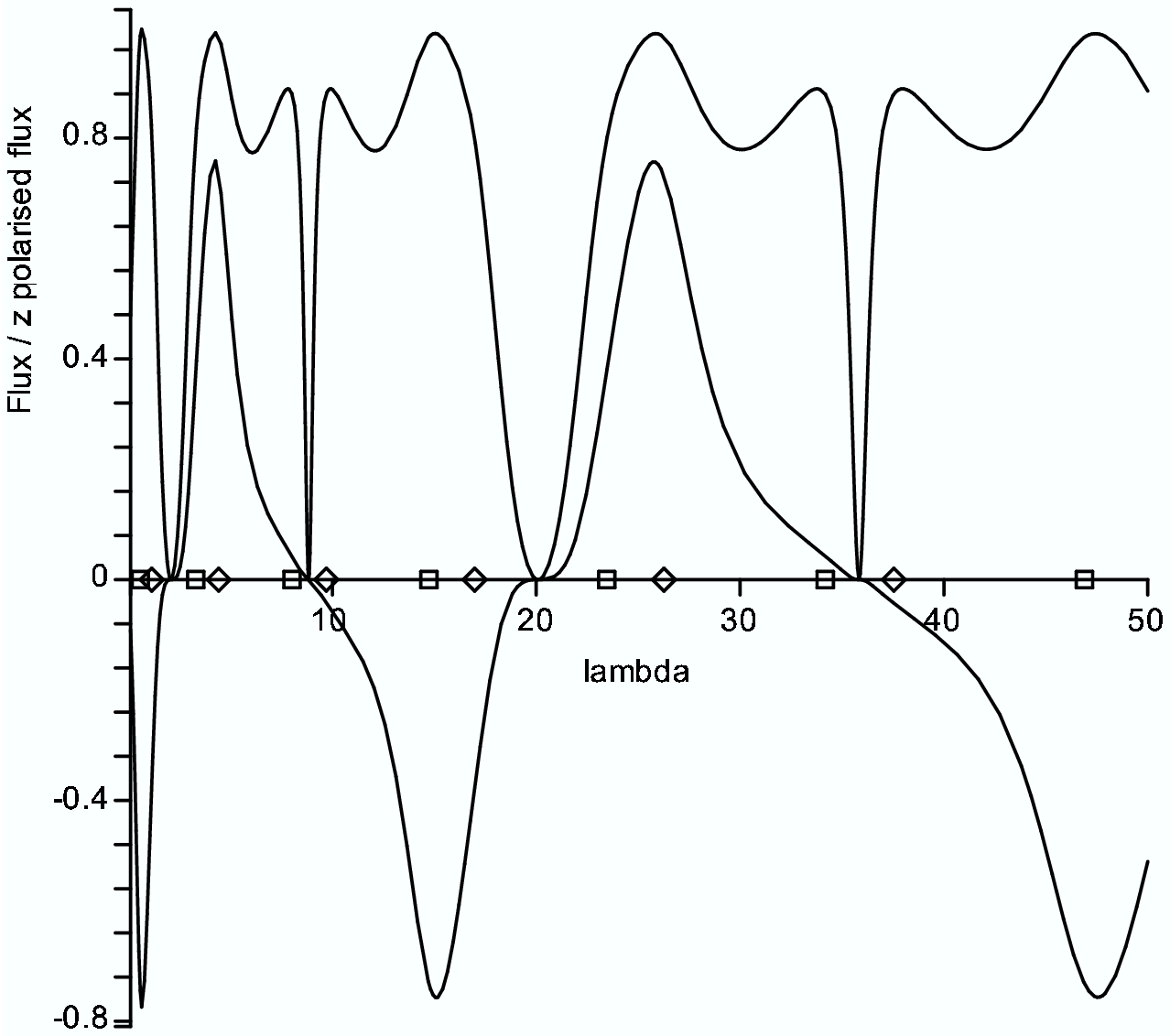}\vspace*{-7.6cm}
\caption{$T_{21}$ and $P_{21,z}$ for $\alpha=0.8$ and $\xi=2\pi/3$.}\label{2pi3}
\end{center}
\end{figure}

\begin{figure}[ht]
\begin{center}
\vspace*{-1.1cm}
\includegraphics[height=15cm,width=14cm]{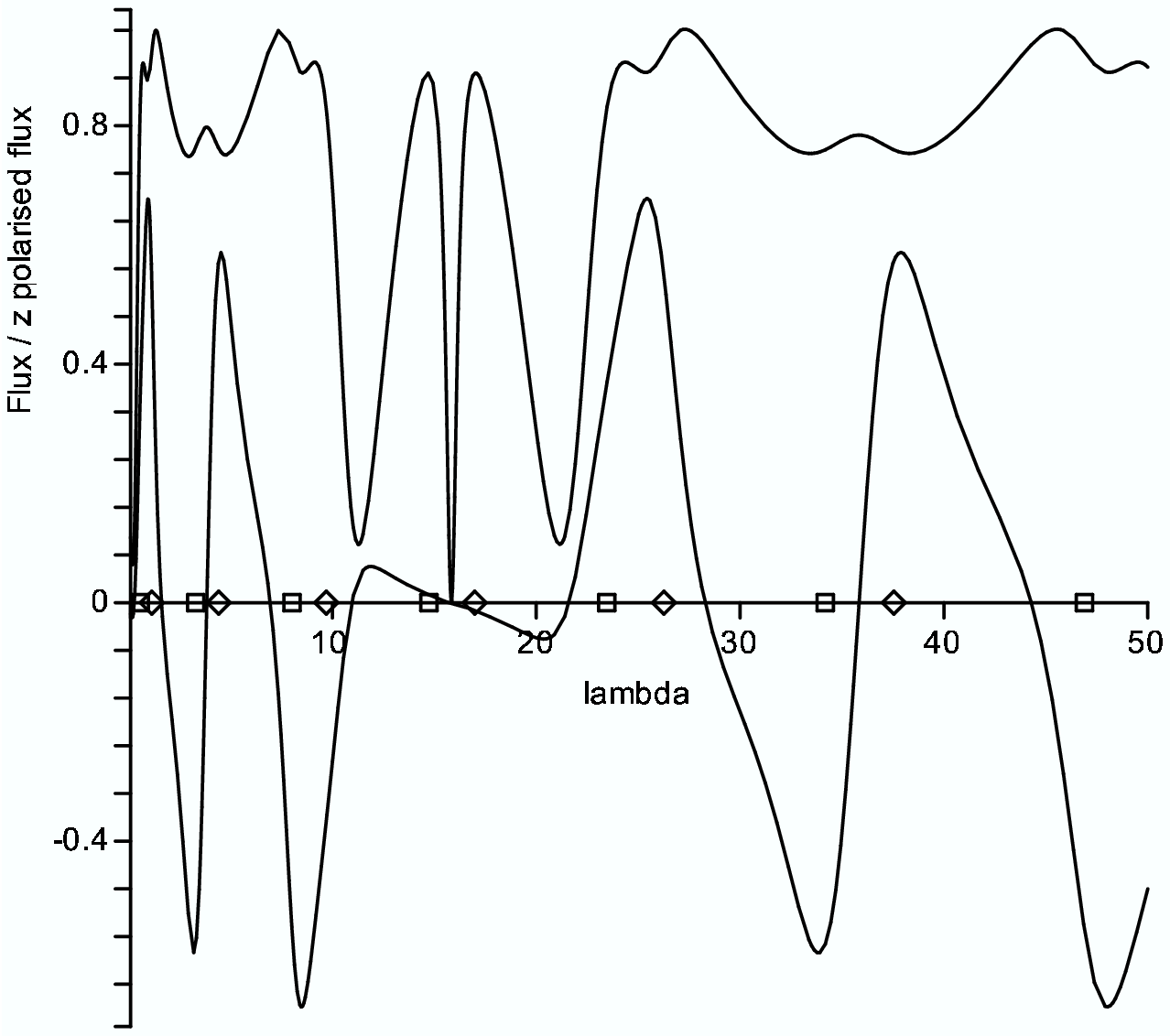}\vspace*{-7.6cm}
\caption{$T_{21}$ and $P_{21,z}$ for $\alpha=0.8$ and $\xi=\pi/4$.}\label{pi4}
\includegraphics[height=15cm,width=14cm]{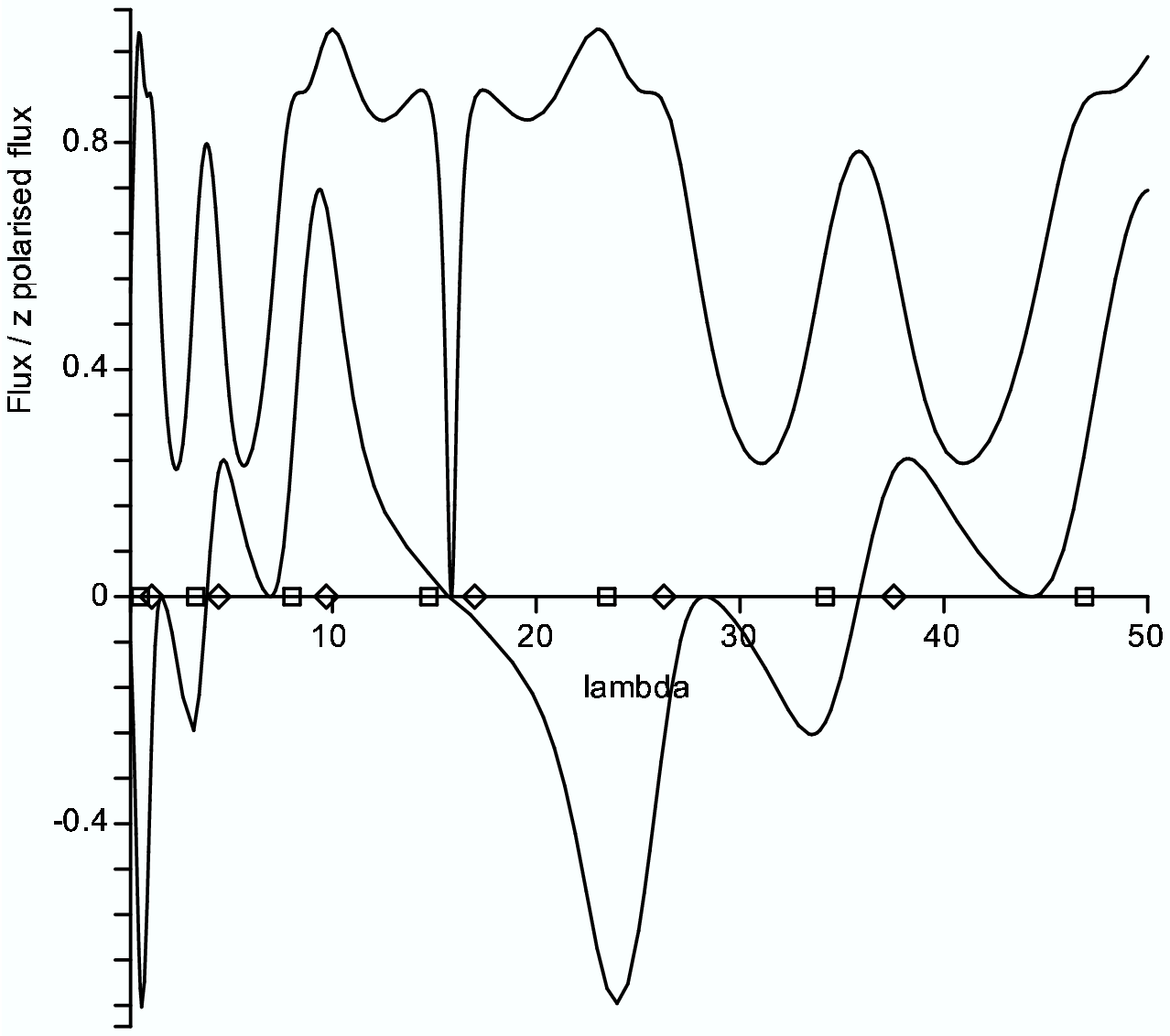}\vspace*{-7.6cm}
\caption{$T_{21}$ and $P_{21,z}$ for $\alpha=0.8$ and $\xi=3\pi/4$.}\label{3pi4}
\end{center}
\end{figure}

Let us consider a a three terminal device with symmetry as illustrated in figure \ref{sfil}, ie. the angle $\xi\in(0,\pi)$ between the first and second and first and third wires the same.
To be precise we also need that the coupling constants at vertices two and three are the same---in fact, for simplicity, we will set all of the coupling constants equal to one, $\beta_i=1$. \\
As was shown in \cite{Kis:Kim1} a three terminal device with this symmetry can potentially act as a spin filter. Specifically, for unpolarised current entering the first wire the polarisation of flux measured on wires two, $P_{21,\alpha}$, and three, $P_{31,\alpha}$, along the $\alpha$-axis satisfies
$$
P_{21,x} = - P_{31,x} \, , \quad P_{21,y} = P_{31,y} \, , \quad P_{21,z} = - P_{31,z} \, .
$$
A proof of these statements, following \cite{Kis:Kim1} but in the context of quantum graphs, is given in appendix 2. \\
Appendix 3 contain an outline of the derivation of expressions for the conductance $T_{21}$ and the polarisation in the $z$-axis $P_{21,z}$ for current going from wire one to two for the device in figure \ref{sfil}. In figures \ref{pi2}--\ref{3pi4} we plot the conductance $T_{21}$ (upper curve) and $P_{21,z}$ (lower curve) against the energy $\lambda=k^2$. In these plots we assume $\alpha=0.8$; there is no significant change in the form of the $T_{21}$ and $P_{21,z}$ curves with respect to $\alpha$ apart from at the special values discussed below. We assume that the angle of attachment of the wires, $\xi=p\pi/q$, is an integer fraction of $\pi$. 
The resonance eigenvalues of the ring are indicated by $\Box$ for $\lambda_{+,n}$ and $\Diamond$ for $\lambda_{-,n}$ (the leftmost $\Box$ and $\Diamond$ are at $\lambda_{+,1}$ and $\lambda_{-,1}$ respectively with $\lambda_{-,0}\le0$ off the left of the plot). \\
It is clear from these plots that at some of the resonant energies, depending on the value of $\xi$, this device would function as an efficient spin filter. Our case $\xi=\pi/2$, figure \ref{pi2}, most closely resembles the device considered in \cite{Kis:Kim1} and we observe as did Kiselev and Kim that the polarisation is maximum at odd resonant eigenvalues and severely damped at even eigenvalues---an explanation in terms of interference effects is given in \cite{Kis:Kim1}. However, for our device at $\xi=\pi/2$ the polarisation changes sign at the odd resonances indicating that this may not be an ideal parameter regime for spin filtering. Clearly defined and isolated peaks can be found for instance at $\xi=2\pi/3$ and we believe would be a better parameter regime for spin filtering. \\ 
We remark that for $\xi=p\pi/q$, $T_{21}$ and $P_{21,z}$ are periodic repeating after $2q$ resonances $\lambda_{\pm,n}$ with zeroes in $T_{21}$ and $P_{21,z}$ at $\lambda_{\pm,kq}$. It seems plausible that this behaviour can again be explained by spin interference and topological phase effects \cite{LyGe}. \\
Finally, we observe from the expression for the polarisation $P_{21,z}$, equation (\ref{fap}), the following behaviour as a function of the Rashba coefficient $\alpha$. As noted above, for  $\sqrt{1+\alpha^2} -1=m\in \{0,1,\ldots\}$ the eigenvalues on the ring become fourfold degenerate. For these values of $\alpha$, $\kappa_{+}=\kappa_{-}+m+1$ so that---see the expression for $Q$ given after equation (\ref{fap})---$Q\equiv 0$ and the polarisation is {\em identically zero}. This is clearly a generalisation of the degenerate behaviour at $\alpha=0$, a possible physical explanation involves noting that there is an integral number of effective flux quanta \cite{SGZ} through the ring for these values of $\alpha$ and showing that for integral quanta the spin interaction is cancelled. We rather consider the gauge transformation 
$$
V H_0 V^{\star} = -\frac{d^2}{d\theta^2} - \left( \frac{\alpha}{2} \right)^2
$$
on the ring where $V (\theta) = e^{-i\sigma_{z} (m+1) \theta/2}\, e^{i\sigma_{y} \varphi/2}\, e^{i\sigma_{z} \theta/2}$. Here, since $V (\theta)$ is not single valued, we take $\theta\in[0,2\pi )$ and assume that the point of attachment $\theta_1 = 0$. We also make a change of basis on the wires, formally $V(\theta_j) H_j V^{\star} (\theta_j)$. Generally this results in all spin interaction being concentrated at the boundary conditions; however, for $\sqrt{1+\alpha^2} -1=m\in \{0,1,\ldots\}$ the boundary conditions remain spin independent so that, in analogy with $\alpha=0$, we have vanishing of the polarisation. Precisely, at all vertices except $\theta_1 =0$ the original boundary conditions (\ref{bndcnd}) continue to hold while at the origin we get the new boundary conditions
$$
\beta^{-1} \psi_1 = \psi_2 = (-1)^{m} \psi_3 \, , \quad \beta\psi^{\prime}_{1} + \psi^{\prime}_{2} + (-1)^{m} \psi^{\prime}_{3} = 0 \, .
$$
Here edge $i=1$ is the semi-infinite wire, edge $i=2$ corresponds to $\theta\in(0,\theta_2)$, edge $i=3$ corresponds to $\theta\in(\theta_n,2\pi)$ on the ring and we recall that the above derivatives are in the outward direction from the vertex. These boundary conditions show that, up to an energy shift on the ring, the scattering properties of the ring fall into two classes depending on whether $m$ is even or odd indicating that the conductance is a periodic function of $\alpha$ (as observed in the case of a two terminal device in \cite{NMT}). 
\section*{Acknowledgements}
The author has benefitted greatly from conversations with Prof B.Pavlov and Dr U. Z\"{u}licke. 
\section*{Appendix: Scattering matrix for the T-junction}
There is a well established description of the scattering matrix for the T-junction in the physics literature (see \cite{Datt}, pg 173, and \cite{Tan:But, SGB}). In this appendix we show how this ans\"{a}tz is related to the solvable models approach of specifying boundary conditions (\ref{bndcnd}) at the vertex of the T-junction. \\
We note that (\ref{bndcnd}) are of `projection type' \cite{Har4}, ie. we can express these boundary conditions in the form 
\begin{equation}\label{bndcnd2}
P^{\perp}\bar{\psi}=0 \, , \quad P\bar{\psi}^{\prime}=0
\end{equation}
where $\bar{\psi}=\left( \psi_1 , \psi_2 , \psi_3 \right)^{T}$, $\bar{\psi}^{\prime}=\left( \psi^{\prime}_1 , \psi^{\prime}_2 , \psi^{\prime}_3 \right)^{T}$ and 
$$
P = \frac{1}{\beta^2 + 2} \left( \begin{array}{ccc}
\beta^2 & \beta & \beta \\
\beta & 1 & 1 \\
\beta & 1 & 1
\end{array} \right) \, ,
$$
$P^{\perp}=\I - P$ are projections. Now we suppose that the T-junction is at the point where three semi-infinite wires meet, instead of at the point of connection of one wire and a ring. As above we construct scattered waves for this non compact system and derive the scattering matrix which turns out to be \cite{Har4}
$$
S = 2P-\I = \frac{1}{\beta^2 + 2} \left( \begin{array}{ccc}
\beta^2 - 2 & 2\beta & 2\beta \\
2\beta & -\beta^2 & 2 \\
2\beta & 2 & -\beta^2
\end{array} \right) \, .
$$
This scattering matrix is the same as posited in \cite{Datt,SGB,Tan:But} for the T-junction (a different ordering of the edges is used in the last two references) motivating our choice of (\ref{bndcnd}) for the boundary conditions. \\
We note that, at least for the chosen boundary conditions (\ref{bndcnd}, \ref{bndcnd2}), scattering at the idealised T-junction is independent of energy. It is not difficult to see that this is true for any projection type boundary condition and furthermore means that there are no discrete eigenvalues or resonances associated to the junction (this is an equivalence \cite{Har4}, there are no discrete eigenvalues or resonances associated to the junction iff the boundary conditions are of projection type). We also note that projection type boundary conditions, also referred to in the degree two case as F\"{u}l\"{o}p-Tsutsui point interactions, are important in the study of `quantum chaotic' behavior in quantum graphs \cite{Hej:Che} due to their scale invariance. \\
Given a two or three dimensional quantum network, the problem of deriving appropriate boundary conditions at the vertices of an approximating one dimensional quantum graph is an active area of research (see \cite{BMPPY, Exn:Nem, HPY}) as is the `inverse problem' of constructing sequences of graphs with regular potentials so that in the limit we observe a chosen boundary condition from the whole $U(n)$ parameter space \cite{ENZ, Che:Exn}.
 \section*{Appendix: Symmetries of the scattering matrix}
Here we follow the argument of \cite{Kis:Kim1, Kis:Kim2}, which describes the symmetries of the scattering matrix, using terms appropriate for quantum graphs. \\
 We think of $\Gr$ as a graph in the plane and suppose that $\gamma$ is a closed curve nowhere tangent to $\Gr$. The wronskian is defined as
$$
W_{\gamma} \left( f , g \right) = \sum_{x_i} (-1)^{\sigma(x_i)} \left. \left( \langle D f , g \rangle + \langle f , D g \rangle \right) \right|_{x_i}
$$
where $\{ x_i \} = \gamma \cap \Gr$ and $\sigma(x_i)$ is the orientation of the ordered pair formed of the orientation of the local variable at $x_i$ and the orientation of $\gamma$ at $x_i$. The operator $D$ is one of $D_0$ or $D_j$ depending on whether $x_i$ is on the ring or the wires. \\
We always assume that the wronskian acts on solutions of the eigenequation, $Hf=k^2 f$, $Hg=k^2 g$, from which it is easy to see that $W_{\gamma} \left( f , g \right)$ is piecewise constant. Furthermore, we assume that $\gamma$ is large, in particular it encloses and has no intersections with the ring, in which case we drop the subscript and write
$$
W \left( f , g \right) = \sum_i \left. \left( \langle D_i f , g \rangle + \langle f , D_i g \rangle \right) \right|_{x_i} \, .
$$
From the constancy of the wronskian we see that on the ring
$$
\sum^n_{j=1} \left. \left( \langle D_0 f , g \rangle + \langle f , D_0 g \rangle \right) \right|^{\theta = \theta^{+}_j}_{\theta = \theta^{-}_j} = 0
$$
while on the wires
$$
\left. \left( \langle D_i f , g \rangle + \langle f , D_i g \rangle \right) \right|_{x_i} = \left. \left( \langle D_i f , g \rangle + \langle f , D_i g \rangle \right) \right|_{x_i = 0}
$$
so that $W \left( f , g \right)$ is actually equal to the boundary form (\ref{bform}). In particular, if $f$ and $g$ are eigensolutions satisfying the boundary conditions, ie. such that the boundary form (\ref{bform}) vanishes, then
$$
W \left( f , g \right) = \left( H f , g \right) - \left( f , H g \right) = 0 \, .
$$
(In fact since our boundary conditions are `local' we will have $W_{\gamma} \left( f , g \right) = 0$ for any $\gamma$, but we do not need this.) \\
The wronskian allows us to identify symmetries of the scattering matrix. Consider the wronskian of two scattered waves:
\begin{eqnarray*}
0 & = & W \left( \psi_{i} , \psi_{j} \right) = \sum_{k} \left. \left( \langle D \psi_{i} , \psi_{j} \rangle + \langle \psi_{i} , D \psi_{j} \rangle \right) \right|_{x_k = 0} \\
& = & \sum_{k} -k \left( \left( \delta_{ik} - S^{\star}_{ik} \right)
\left( \delta_{kj} + S_{kj} \right) + \left( \delta_{ik} + S^{\star}_{ik} \right)
\left( \delta_{kj} - S_{kj} \right) \right) \, ,
\end{eqnarray*}
we get immediately
$$
S^{\star} S = \I \, .
$$
We note that in this case, since $\psi_{i}$ is matrix valued, the wronskian $W \left( \psi_{i} , \psi_{j} \right)$ is properly thought of as a $2\times 2$ matrix. \\
Further symmetries of the scattering matrix may be found from operators commuting with the hamiltonian. Here we are mainly interested in the case where the graph, and boundary conditions, are invariant with respect to a reflection in one of the coordinate axes $R: y\leftrightarrow -y$. It is then clear that the hamiltonian will commute with ${\cal R} = \sigma_{y} R$ and we have the vanishing of the wronskian
\begin{eqnarray*}
0 & = & W \left( {\cal R} \psi_{i} , \psi_{j} \right) = \sum_{k} \left. \left( \langle D {\cal R} \psi_{i} , \psi_{j} \rangle + \langle {\cal R} \psi_{i} , D \psi_{j} \rangle \right) \right|_{x_k = 0} \\
& = & \sum_{k} k \left( \left( \delta_{ik} \sigma_{y} - R(S^{\star}_{ik}) \sigma_{y} \right) \left( \delta_{kj} + S_{kj} \right) - \left( \delta_{ik} \sigma_{y} + R(S^{\star}_{ik}) \sigma_{y} \right)
\left( \delta_{kj} - S_{kj} \right) \right)
\end{eqnarray*}
or
\begin{equation}\label{Rsym}
 \hat{\sigma}_{y} R ( S^{\star} ) \hat{\sigma}_{y} S = \I \; \Rightarrow \; S = \hat{\sigma}_{y} R \left( S \right) \hat{\sigma}_{y} \, .
\end{equation}
Here $\hat{\sigma}_{y}$ is block diagonal with $\sigma_{y}$ on the diagonal. \\
Before we apply this we need to define some notation. It is convenient for us to decompose the components of the scattering matrix in terms of spin matrices 
$$
s_{ij} = s_{ij,1} + i \sum_{\alpha} \sigma_{\alpha} s_{ij,\alpha} \, .
$$
In terms of this decomposition we can express the conductance $T_{ij}$ and the polarisation in the $\alpha$-axis $P_{ij,\alpha}$ for waves going from wire $j$ to wire $i$ as
\begin{eqnarray}
T_{ij} & = & 2 \left( \left| s_{ij,1} \right|^2 + \left| s_{ij,x} \right|^2 + \left| s_{ij,y} \right|^2 + \left| s_{ij,z} \right|^2 \right) \label{flux} \\
P_{ij,\alpha} & = & 4 \Im \left( s_{ij,1} \bar{s}_{ij,\alpha} + s_{ij,\alpha-1} \bar{s}_{ij,\alpha+1} \right) \, . \label{pol}
\end{eqnarray}
Now we consider the three terminal device illustrated in figure \ref{sfil} which is clearly invariant with respect to $R$. Then (\ref{Rsym}) implies 
\begin{eqnarray*}
& s_{ij,1} = s_{i'j',1} \, , \quad s_{ij,y} = s_{i'j',y} & \\
& s_{ij,x} = -s_{i'j',x} \, , \quad s_{ij,z} = -s_{i'j',z} &
\end{eqnarray*}
where $R (s_{ij}) = s_{i'j'}$. In particular, the polarisation satisfies
\begin{eqnarray*}
& P_{21,x} = 4 \Im \left( s_{21,1} \bar{s}_{21,x} + s_{21,z} \bar{s}_{21,y} \right) = - P_{31,x} & \\
& P_{21,y} = 4 \Im \left( s_{21,1} \bar{s}_{21,y} + s_{21,x} \bar{s}_{21,z} \right) = P_{31,y} & \\
& P_{21,z} = 4 \Im \left( s_{21,1} \bar{s}_{21,z} + s_{21,y} \bar{s}_{21,x} \right) = - P_{31,z} & \, .
\end{eqnarray*} 
Finally we note that for the system under consideration the time reversal operator ${\cal K} = \sigma_{y} K$, where $K$ is complex conjugation, along with ${\cal L} = \sigma_{z} L$, where $L$ reverses the sign of $\alpha$ or equivalently $\varphi$, both commute with the hamiltonian. Proceeding as above these can be used to find yet more symmetries of the scattering matrix (see \cite{Kis:Kim2} for a further discussion).
\section*{Appendix: Derivation of polarisation and conductance}
Using equations (\ref{Gfn}, \ref{Smatrix}) we see that we can express the scattering matrix for the device illustrated in figure \ref{sfil} as
$$
S = U^{\star} \left( \begin{array}{ccc}
\bar{b} & \bar{z}_{1} & z_{1} \\
z_{1} & \bar{b} &  z_{2} \\
\bar{z}_{1} & \bar{z}_{2} & \bar{b}
\end{array}
\right) \left( \begin{array}{ccc}
b & \bar{z}_{1} & z_{1} \\
z_{1} & b &  z_{2} \\
\bar{z}_{1} & \bar{z}_{2} & b
\end{array}
\right)^{-1} U
$$
where 
\begin{eqnarray*}
U & = & e^{i\sigma_{y}\varphi/2}\left( \begin{array}{ccc}
1 & 0 & 0 \\
0 & e^{j\xi/2} & 0 \\
0 & 0 & e^{-j\xi/2}
\end{array}
\right) \, , \\
z_{l} & = & j \kappa \left( \frac{e^{j\kappa_{+}(l\xi -\pi)}}{\cos(\kappa_{+}\pi)} - \frac{e^{-j\kappa_{-}(l\xi -\pi)}}{\cos(\kappa_{-}\pi)}\right) \, , \\
b & = & z_{0} + i \beta^{-2} = \kappa \left( \tan(\kappa_{+}\pi) + \tan(\kappa_{-}\pi) \right) + i \beta^{-2} \, ,
\end{eqnarray*}
$\kappa = -k/2(\kappa_{+} + \kappa_{-})$, $j=i\sigma_{z}$ and we have assumed that all of the coupling constants are equal, $\beta_i =\beta$. \\
From the form of the scattering matrix we see that 
$$
s_{21} =  e^{-j\xi/2} e^{-i\sigma_{y}\varphi/2} \left( {\sf s}_1 + i \sigma_{z} {\sf s}_z \right) e^{i\sigma_{y}\varphi/2}
$$
where, using Maple, 
\begin{equation}\label{s1sz}
{\sf s}_1 + j {\sf s}_z = \frac{b-\bar{b}}{D} \left( z_{1} b - \bar{z}_{1} z_{2} \right)
\end{equation}
and, due to the form of $z_l$, the determinant 
$$
D = b^3 - b z_2 \bar{z}_2 - 2 b z_1 \bar{z}_1 + z^2_1 \bar{z}_2 + \bar{z}^2_1 z_2
$$
is a complex scalar. \\
It is easy to show that the terms due to $U$ in the expression for $s_{21}$ make no contribution to the conductance, equation (\ref{flux}), 
$$
T_{21} = 2 \left( \left| {\sf s}_{1} \right|^2 + \left| {\sf s}_{z} \right|^2 \right)
$$
and introduce a multiplicative factor into the polarisation in the $z$-axis, equation (\ref{pol}),
$$
P_{21,z} = 2i \cos(\varphi) \left( \bar{{\sf s}}_{1} {\sf s}_{z} - {\sf s}_{1} \bar{{\sf s}}_{z} \right) \, .
$$
Using (\ref{s1sz}) we write ${\sf s}_{1}$ and ${\sf s}_{z}$ in terms of $z_{l}$ and $b$ (here again the form of $z_{l}$ is important) which gives us
$$
T_{21} = \frac{ 8 }{\left| D \right|^2} \left( \left( | b |^2 + | z_2 |^2 \right) | z_1 |^2 - {\textstyle \frac{1}{2}} \left(b + \bar{b} \right) \left( \bar{z}^2_1 z_2 + z^2_1 \bar{z}_2 \right) \right)
$$
and 
$$
P_{21,z} = \frac{8 \cos(\varphi)}{\left| D \right|^2}\, j \left( \bar{z}^2_1 z_2 - z^2_1 \bar{z}_2 \right) \, .
$$
We again use Maple to get explicit expressions and simplify---here it is important to cancel common factors $\cos^{-2}(\kappa_{+}\pi)\cos^{-2}(\kappa_{-}\pi)$ which appear in the numerator and denominator to aid simplification and avoid numerical instability. At this step we also put $\beta=1$. The expression for the conductance and polarisation are then
\begin{equation}\label{fap}
T_{21} (k,\xi,\alpha) = \frac{8 R}{X^2 + Y^2} \, , \quad P_{21,z} (k,\xi,\alpha) = \frac{8 \cos(\varphi)\, Q}{X^2 + Y^2} \, ,
\end{equation}
where 
\begin{eqnarray*}
X & = & \mbox{} - \left( 4 \kappa^3 + 3 \kappa \right) \sin \left( \kappa_{+} + \kappa_{-} \right) \pi 
+ 4 \kappa^3 \sin \left( \kappa_{+} + \kappa_{-} \right) \left( 2\xi-\pi \right) \\
& & \mbox{} - 8 \kappa^3 \sin \left( \kappa_{+} + \kappa_{-} \right) \left( \xi-\pi \right) \\
Y & = & \mbox{} - \left( 6 \kappa^2 + {\textstyle \frac{1}{2}} \right) \cos \left( \kappa_{+} + \kappa_{-} \right) \pi - {\textstyle \frac{1}{2}} \cos \left( \kappa_{+} - \kappa_{-} \right) \pi \\
& & \mbox{} + 2 \kappa^2 \cos \left( \kappa_{+} + \kappa_{-} \right) \left( 2\xi-\pi \right) + 4 \kappa^2 \cos \left( \kappa_{+} + \kappa_{-} \right) \left( \xi-\pi \right) \\
R & = & \kappa^2 + 4 \kappa^4 + {\textstyle \frac{1}{2}} \kappa^2 \left[ \cos\left(2 \kappa_{+}\pi\right) + \cos\left(2 \kappa_{-}\pi\right) - \cos \left( \kappa_{+} + \kappa_{-} \right) \xi \right. \\
& & \mbox{} - \cos \left( \kappa_{+} + \kappa_{-} \right) \left( \xi-2\pi \right) - \cos \left( \kappa_{+} \left( \xi-2\pi \right) + \kappa_{-} \xi \right) \\
& & \left. \mbox{} - \cos \left( \kappa_{+} \xi + \kappa_{-} \left( \xi-2\pi \right) \right) \right] \\
& & \mbox{} + 2 \kappa^4 \left[ \cos \left( \kappa_{+} + \kappa_{-} \right) \left( \xi-2\pi \right) + \cos \left( \kappa_{+} + \kappa_{-} \right) \left( 3\xi-2\pi \right) \right] \\
& & \mbox{} + 4 \kappa^4 \left[ \cos \left( \kappa_{+} + \kappa_{-} \right) \xi + \cos \left( \kappa_{+} + \kappa_{-} \right) 2 \left( \xi-\pi \right) \right] \\
Q & = & \kappa^3 \left[ \cos\left(2 \kappa_{-}\pi\right) - \cos\left(2 \kappa_{+}\pi\right) \right. \\
& & \left. \mbox{} + \cos\left(  \kappa_{+}2\xi + \kappa_{-} 2\left(\xi - \pi \right) \right) - \cos\left( \kappa_{+} 2\left(\xi - \pi\right) + \kappa_{-}2\xi \right) \right] \\
& & \mbox{}+ 2 \kappa^3 \left[ \cos\left( \kappa_{+} \left(\xi - 2\pi\right) + \kappa_{-}\xi \right) - \cos\left(  \kappa_{+}\xi + \kappa_{-} \left(\xi - 2\pi\right) \right) \right] \, .
\end{eqnarray*}

\end{document}